# Unlikely Organizers:
# The Rise of Labor Activism Among Professionals in the U.S. Technology Industry

JS Tan
Massachusetts Institute of Technology
77 Massachusetts Ave
Cambridge, MA, US 02139-4307
js_tan@mit.edu
+1 (617) 888-3552
Corresponding author
ORCiD ID: 0009-0001-5864-9721

Natalia Luka
University of California, Berkeley
410 Social Sciences Building
Berkeley, CA, US 94720
hello@natalialuka.com
+1 (732) 616-4909
ORCiD ID: 0000-0003-1576-7346

Emily Mazo
Columbia University
606 W 122nd St, Floor 5
New York, NY, US 10027-6902
erm2192@columbia.edu
+1 (267) 975-1578
ORCiD ID: 0009-0001-6368-3641

Acknowledgments: We thank Jamie Woodcock, Adam Reich, Susan Silbey, Jason Jackson, Heather Haveman, and Neil Fligstein for feedback on earlier versions of this paper. We also benefited from comments by participants in the 2024 American Sociological Association conference and the Center for Culture, Organizations, and Politics workshop at UC Berkeley. Wynnie Chan, Hyatt Dirbas, and Claire Ma made substantial contributions to the dataset for this research project. All errors are our own.

Note: Data is available at https://data.collectiveaction.tech. Additional results and programs used to generate the results presented in the article are available from the corresponding author.

In 2018, at the height of the #metoo movement, the New York Times broke a story about a $90 million settlement paid to Andy Rubin, a former Google executive who was asked to resign after credible claims of sexual harassment (Wakabayashi et al. 2018). The revelation formed the basis for a walkout of 20,000 Google employees, many of whom would eventually join the company's minority union, the Alphabet Workers' Union. While the spotlight of this protest was on discrimination against women and people of color, the demonstration also featured calls for improved working conditions for all Google employees.

In recent years, tech professionals—including software engineers, product managers, UX designers, etc.—have started participating in labor activism. Google contractors at HCL Technologies won the first-ever union of software engineers in the US in 2019, followed shortly by Kickstarter employees who formed the first wall-to-wall union in the tech sector. In 2022, hundreds of tech workers at the New York Times voted 404 to 88 to certify their union. Seeing this opportunity, national unions, including the Communication Workers of America (CWA), United Steel Workers, and the Office and Professional Employees International Union (OPEIU) have also poured resources into organizing tech professionals. At the time of publication, there are 20 recognized unions in the industry.

Conventionally, scholars have understood labor activism as a response to exploitation and a means to improve economic and working conditions. Yet tech professionals are well compensated, afforded considerable workplace autonomy, and have high labor market power, allowing them to switch jobs rather than organize for change from within the firm. Moreover, they have historically self-identified as professionals, creatives, or entrepreneurs rather than workers (Dorschel 2022a; Rothstein 2022), making them conventionally seen as incompatible with labor activism (Dorschel 2022b, Van Jaarsveld 2004; Milton 2003; Brophy 2006). So, why did tech professionals start engaging in labor activism and participating in the American labor movement?

To explain the rise of labor activism in the tech industry, we collected and analyzed all publicly documented instances of collective action by tech workers from 2014 to 2022. Over this period, tech professionals engage not only in labor activism but in another type of workplace activism that we call social activism, activism over issues that do not directly impact the participating employees. Employee protests against military contracts, for example, fit the bill since potential harm caused by the military does not directly affect the employees protesting the contract. Our data analysis reveals that a period of social activism preceded the labor activism that is now prevalent across the industry. Regression analysis of our dataset confirms that participation in social activism has an effect on the likelihood of labor activism.

Tech workers possess a distinctive professional identity centered on using technology for societal benefit. However, when employers undermine these idealistic values by engaging with perceived bad actors, such as oil companies or militaries, they violate tech workers' belief in contributing to the social good. In response, tech workers, guided by their unique professional identity in preserving the social good, engage in social activism in the workplace. Participation in such activism generates solidarity among employee-participants. It also creates conflict with management, who sees such protest strategies as tarnishing the company's brand image. In this way, social activism exposes the previously hidden divide between workers and management, masked by the industry's techno-utopianism. Consequently, employees come to understand their own identities not just as professionals but as workers, creating the basis for labor activism.

Our analysis comprises three parts. First, we conducted an analysis of the two types of workplace activism within the tech sector, finding evidence of a shift from social to labor activism across the industry. Second, using a logistic regression model and company fixed effects, we test the relationship between the two types of workplace activism and the effects that organizational types have on labor activism. Finally, to explore the mechanism in which labor activism emerges, we conduct a case study analysis of workplace activism at Google.

## Do professionals participate in labor activism?

Professionals can be defined as workers with jurisdiction over a particular set of skills or decisions for which a labor market exists. In academic literature, professionals are portrayed at the top of an ecosystem of employment in the U.S., protected by high barriers to entry for their profession, allowing them to be able to negotiate for high compensation and a degree of self-governance through professional organizations (Abbott 1988; Freidson 2001; Brint 1996). From their advantaged position, professionals would seem to have neither the need nor the interest to challenge existing systems of power in the workplace.

Empirical accounts confirm these challenges and demonstrate how workers use professional ideologies to justify their lack of involvement in labor activism. Ideologies can exist at the level of the profession but can also permeate entire industries associated with professionalism, such as tech or finance. In their account of hospital worker organizing, Fink and Greenberg (2009) note the difficulty of getting pharmacists to strike and unionize alongside service and maintenance workers due to "reservations about mixing 'professionalism' and 'unionism'" (116). Within the same workplace, inequalities in pay, treatment, and respect can allow one set of workers to distance themselves from the struggles of others. In other cases, entire industries can be infused with an anti-labor sentiment based on a more general ideology of professionalism. In a case highly relevant to our own, Milton describes the challenges of unionizing "high-tech talent," arguing that workers saw unions as "opposed to the

meritocracy…that anchors excellence in technology-based industries" (2003, 41). The workers Milton describes encompass numerous different professions and span several companies, but all identify with an industry-wide neoliberal ideology that Barbrook and Cameron summarize as a "[r]elentless drive to…intensify the creative powers of human labor" (1996, 44). Dorschel (2021) and English-Lueck (2011) have noted similar meritocratic and entrepreneurial ideologies among tech workers, suggesting that these professionals face a two-fold challenge of combating both the ideologies of their individual professions and the greater industry.

Yet instances of labor activism among professionals do exist. Teachers and nurses unions have been among the most significant groups in organized labor in the U.S. in the past few decades (Weiner 2012; Dube, Kaplan, and Thompson 2016). One explanation for higher rates of labor activism by teachers and nurses is that these professions have been historically underpaid, more likely to be staffed by women and people of color than other professions, and only recently seen as "professional," partly as a result of activism that advocated for the importance for this work (Apesoa-Varano and Varano 2004; Fink and Greenberg 2009). Journalists, another set of professional workers who have also engaged in a significant amount of labor activism, were, in the mid-century, in part influenced and incited to unionize when they saw blue-collar print workers receiving higher wages as a result of their success in organizing (Salamon 2023). They, too, underwent a gradual professionalization process. In recent years, the profession has been hit with a rise in precarity and the proliferation of freelance journalistic work, leading to a resurgence of labor activism since 2015 (Salamon 2020; Luce and Shaffer 2020). Similarly, the recent unionization successes of medical residents were largely in response to the COVID-19 pandemic and increased burnout and dissatisfaction over long hours and low pay (Lin, Ge, and Pal 2022).

What all these groups have in common is a degree of precarity and exploitation despite their professional or professional-in-training status. Still, labor activism among most professional categories remains rare, and tech workers represent perhaps the most standout category of labor organizing among professionals in the last decade because of their reported high rates of pay and significant work benefits (Kochan et al. 2023). What stands out, too, about the case of tech workers is that they self-describe their work as creative, innovative, and even entrepreneurial (Thompson 2019). Additionally, tech workers are morally motivated and hold techno-utopian beliefs that technology expands democracy and reduces hierarchy (Turner 2006) while also embracing the wealth creation enabled by new network technologies that can connect individuals in virtual communities—described by Barbrook and Cameron as the 'Californian Ideology' (1995). As such, many high-tech workers select their profession with the belief that their work will be helpful to society (McAdam et al. 2022; Wilmers and Zhang 2022; Augustine and King 2022).

In this way, tech workers, compared to other professional workers, are unique in how they use their careers or professions as platforms for expressing social or political beliefs. This feature of tech workers overlaps with existing literature on occupational activism (Rheinhardt, Briscoe, and Joshi 2023; Coley and Schachle 2023), defined as a type of social activism that relies on ties to an occupational identity or community (Cornfield et al. 2018).[1] To date, there is little documentation of social or occupational activism shifting into labor activism or vice versa, particularly among professionals. Indeed, Coley and Schachle (2023) point to this as an area for further research. In this way, this article presents a rare example that sheds light on the interactions between these two types of workplace activism among professionals. It asks why tech workers—with their moral virtues, individualism, and alignment with employers—also started to participate in labor organizing. It is to this question that we turn next.

---

[1] Occupational activism need not necessarily take place within the workplace, though in practice, it often does.

## Possible explanations behind the rise of labor activism in tech

One explanation for the rise of labor activism among tech professionals is that a shock to labor relations caused workers to respond with labor activism. The economic downturn following the post-pandemic interest rate hike in 2022 caused employers to respond with austerity measures, notably mass layoffs in the tech sector for the first time since the burst of the dot-com bubble (Mickle, Weise, and Grant 2023). Employers have also implemented new practices to tighten control over workers, revoking work-from-home policies, rolling back office perks, and creating new reporting requirements. Literature in labor relations has indeed documented that dissatisfaction with pay, promotion, job security, and working conditions is strongly tied to labor organizing (Schriesheim 1978). This pattern can be traced even among high-tech workers. Rothstein (2022) demonstrates how workers in two semiconductor manufacturing companies were able to contest liberalization narratives from management by creating their own discourse and mobilizing workers' councils in the wake of layoffs. Brophy (2006) shows how contractors at Microsoft seized a political opportunity created by a new labor policy and an existing class-action lawsuit to launch a new union that would fight for improved working conditions.

Sheehan and Williams (2023) follow this tradition and posit that the rise of "bread-and-butter" activism is due to this financial dip, arguing that management used the macro-economic climate as an opportunity to reset labor relations in their favor, precipitating a labor response from workers. However, the data that we present by and large does not support theories of a "shock" in labor relations in 2022, the year industry analysts identify as the start of "tech austerity", or even in earlier years (Mickle, Weise, and Grant 2023). Our data shows a steady rise in labor activism as early as 2019—three years before the resetting of labor relations that Sheehan and Williams refer to. A true "shock" should be experienced as a sudden upsurge in labor activism events, not as a gradual increase beginning years prior to austerity measures. Moreover, many labor actions, such as winning a union election, typically take years to accomplish.[2] This means that labor activism in 2022—the year attributed to the shock in labor relations—would've started long before. If we are attempting to identify external shocks in labor relations that led to the rise of labor activism, we need to look elsewhere.

A second explanation is that external social factors and the political climate led to the rise of labor activism in the U.S. technology industry. Journalistic accounts from the early years of workplace social activism identify Trump's presidential election as an instigating force for a more politically aware and active workforce (Tiku 2019; Tarnoff 2020). As studies have shown, professional tech worker populations in the U.S. are politically left-leaning on the whole and identify strongly with a prosocial vision of the tech industry (Rider 2022; Selling and Strimling 2023). Tiku (2019), in particular, argues that the Trump presidency not only created political events that inspired employee protest but also made workers aware of who in the workplace held similar political views when they attended protests together or signed onto the same open letters, building social ties and bolstering confidence in future protests (Wagner-Pacifici and Tavory 2017). She points to events such as the Muslim travel ban, which incited a number of Google employees to publicly protest at the San Francisco International Airport, notably among them, Google's former CEO Sergey Brin.

---

[2] In-depth single-company case studies of union organizing show that unionizing in tech can take years (Redwine, 2020).

Even if not due to Trump himself, the other social movements and current events that occurred in these years, such as the BLM protests in the summer of 2020, the COVID-19 pandemic (which temporally aligned with the BLM protests of 2020), and a well-publicized series of union drives at Starbucks, Amazon, and elsewhere, could have had an effect on workers deciding to organize labor actions as well as social activism. During this period, political polls showed growing support for labor among left-leaning populations in the U.S. (Saad 2023). Labor, in other words, became one of the issues most fervently adopted by the political left in the U.S., and the technology industry– as one of the industries with the highest proportion of left-leaning workers– saw more labor activism as workers took up these causes. For these reasons, we might expect to see a positive relationship between the incidence of social activism in the workplace and particular historical events.

The political zeitgeist is undoubtedly important, but it alone does not explain why tech workers—and not other left-leaning, high-paid professionals—participate in labor activism. If it did, we would expect professionals in law, structural engineering, architecture, or clinical research to also participate in labor activism.[3] Yet, there is no such evidence. Moreover, pegging the shift to Trump's election would suggest a significant dip in employee activism under the Biden administration, for which there is also no evidence. Instead, our data shows a steady rise in tech worker's participation in labor activism starting in the final years of Trump's presidency and *before* the BLM protests, the COVID-19 pandemic, and the nation-wide surge in labor activism. This does not exclude that these events could have prolonged or increased this rise.

While the preceding explanation doesn't give us a clear reason for the surge in labor activism in the tech sector, it does provide a possible cause for social activism in the workplace.

---

[3] In 2022, Verdant Labs made an infographic that shows how people's professions predict their politics based on political donations. See: https://verdantlabs.com/politics_of_professions/

Many early protests by tech professionals directly responded to Trump's presidency and the policies he advocated. This leads us to a third explanation: participation in social activism spurred those same workers to engage in later labor activism. Theorists of class consciousness have modeled the resurgence of labor activism among subsets of U.S. workers using a dual conception of structure and agency (Sewell 1992). The rise of unionizing efforts has been seen as an interplay between structural forces– political movements, industrial conflicts, management strategies– and worker agency– social and ethnic identities, organizational strength, and mobilization of political resources (Fantasia 1988; Delgado 1993). According to Fantasia (1988), "Collective action…was a complex set of interactions in which structured activity…drew on available social-psychological weaponry as well as on more obvious resources (pre-existing networks, work-group culture, familial support) in developing nascent organizational forms" (235). We adopt a similarly dualistic conception of the structural forces and worker agency.

  Existing literature provides us with a clear hypothesis to test: social activism in the workplace draws on worker identity and conceptions of morality but also utilizes the opportunities afforded by particular political platforms and historical moments (Rheinhardt, Briscoe, and Joshi 2023). It can also precipitate conflict between workers and managers at the individual level when workers are demoted or fired for their activism and at the organizational level when demands are not met (Rheinhardt, Poskanzer, and Briscoe 2023). In this way, workplace social activism can lead to the sort of generative conflict that Fantasia (1988) sees as essential for the development of class consciousness. We begin with the most general premise– that social activism, due to its conflictual nature and its propensity for generating solidarity, creates the conditions for professionals to develop class consciousness. For these reasons, we might expect to see a positive relationship between heightened workplace social activism and heightened labor activism:

> *H1: The likelihood of professional tech workers engaging in labor activism at a company will increase if there was workplace social activism at the same company in the preceding time period.*

Within this explanation, we can examine an additional mechanism in greater detail to account for exactly *why* social activism in the workplace sets the stage for labor activism. It is possible that social activism in the workplace itself directly contributed to later labor activism by increasing the organizational strength of professional workers. Organizational strength can best be understood as the social ties that build into sustainable organizing resources, which include informal collectives, formal institutions, or common processes and strategies (McAdam 1982). We posit that organizing social activism in the workplace allows workers to build organizational strength by connecting with other like-minded workers and building sustainable institutions and processes that they return to for future organizing efforts. As our own data shows, protest actions in the U.S. tech industry were organized by a range of different groups, ranging from dispersed ad-hoc groups to short-term collectives that disbanded soon after the event to longer-lasting worker organizations. Each has varying capacities for building trust among workers, creating sustainable and committed constituencies for activism, and building toward labor activism. Thus, we might expect to see a positive relationship between the presence of worker groups and heightened labor activism in the U.S. technology industry.

> *H2: The likelihood of professional tech workers engaging in labor activism at a company will increase if social activism was organized by organized worker groups (unions and employee activism groups) at the same company in the preceding time period.*

In sum, we propose that this third explanation—that social activism leads to labor activism and that social activism produces the organizational strength necessary for sustained labor activism—merits further research. The rest of this article systematically examines the relationship between social and labor activism and how the former led to the latter in the tech sector.

## Research Methods

**Data**

The basis of our analysis is an original dataset[4] of collective actions in the U.S. technology industry from 2014 to 2022, which we gathered using NexisUni news archives. To compile this dataset, we searched for articles where collective action terms[5] occurred within 25 words of employment terms[6] for the computing and information technology industry. To qualify for our archive, events must be "collective" and present "evidence of action" by currently or recently employed "tech workers." Our archive of events is heavily skewed towards actions reported in the English language press because we conducted our search only in English. Moreover, we exclude actions that took place across multiple companies, such as the Tech Against Fascism protests, where engineers, researchers, analysts, and designers across India's and the United States' technology sector signed an open letter condemning the Indian government's repression of citizens during the 2019 democracy uprising. The majority of collective actions in the archive took place at large tech companies such as Google, Microsoft, and Amazon. Together, these three companies amounted to 51 percent of the total recorded collective actions in this study. One reason for this may be because the media biases larger tech companies in their reporting, which would make large tech firms appear with a higher likelihood in our dataset.

---

[4] We made the dataset available online at Collective Action in Tech's Website: https://data.collectiveaction.tech/.
[5] Collective action terms include protest*, petition*, strike*, open letter*, walk out*, union*, boycott*, letter*, lawsuit*, discuss*, and negotiat*.
[6] Employment terms include employee*, worker*, contract*, and labor*.

We filtered our search to only include actions that were organized by or included tech workers in the U.S, resulting in 175 instances of collective action by U.S. based tech workers between 2014 and 2022, inclusive. This dataset also served as the basis for the panel dataset we constructed for our regression analysis, which we describe later.

We find that since 2017, the number of collective actions by tech workers in the U.S. has grown exponentially. In 2017, there were four reported actions in the industry—at Palantir, Google, and Facebook—all protesting the same set of ethical issues—Trump's Muslim registry and immigration ban and the tech industry's involvement and complicity. As illustrated in Figure 1, collective action taken by tech workers exploded in 2018, peaking in 2019, and has remained relatively high in the following years. In 2019 and 2021, the top years for protests, there were 46 in each year. In 2022, there were 28 actions.

**Measures**

Since we're concerned with explaining the rise of labor activism (H1), we differentiated each collective action in the dataset as either workplace *social activism* or *labor activism* based on the nature of the issue being protested. Actions categorized as social activism are those where the cause of harm occurred externally to those participating in the action. These actions call for change outside the workplace. Generally speaking, workers participating in social activism are not directly impacted by the issues they are protesting. Protests against military contracts fit the bill since the opposition is against potential harm caused by the military around the globe. Tech workers at Facebook supporting their cafeteria workers' union drive would also be considered social activism from the perspective of the former group of workers because the union drive they are backing is external to their own working conditions, making their support an effort to serve the social good (Russell 2019).

On the other hand, actions categorized as labor activism are motivated by issues within the workplace of the employees participating in the action. They call for change within the workplace so that the primary impact of a successful protest will be on the workers themselves. This includes actions that target pay transparency, discrimination, or injustices related to employee working conditions. Union campaigns are also considered labor activism. Each entry was reviewed by two individuals—the initial coder and a separate reviewer—to ensure intercoder reliability across entries, and all conflicts were resolved prior to inclusion in the archive.

In this article, we label actions regarding discrimination, harassment, diversity, equity, and inclusion as labor activism unless an explicit connection is made to systemic injustice or a broader social movement, such as Black Lives Matter or #metoo. Take the case of Google's walkout in 2018. To protest against management and demand more executive accountability for their handling of sexual harassment cases, thousands of Google employees participated in a walkout across Google offices around the world (Wakabayashi et al. 2018). Without context, it may seem like this type of collective action should be categorized as labor activism since many of the demands made by protestors are concerned with internal company policy. Moreover, the issue of sexual harassment from an executive is an abuse of managerial power. However, the walkout took place just one year after the #metoo movement in 2017, which turned sexual harassment and abuse into a social and political movement. We categorized this event as both social and labor activism because, while workers were the recipients of harm perpetuated by the target of the protest in this case, the event organizers also referenced the protest as one of a series of fights in a broader systemic struggle against sexism and disempowerment in the workplace. Thus, while the topic of protest is considered "bread-and-butter," the motivation behind the action is aligned with the moral values of the tech worker-professional identity.

Unions frequently engage in protest action of the type we categorize in this dataset as "social activism" either through frameworks of "bargaining for the common good" or "social justice unionism." For example, the Alphabet Workers Union in 2022 partnered with workers at Amazon to write an open letter protesting Project Nimbus, a contract both companies signed with the Israeli military, which we categorized as social activism. We could use terms like "social justice unionism" and "bread and butter unionism," but choose not to because, while the workers engaging in social activism are doing what we recognize as labor organizing, we highlight with our terminology that these workers did not define it that way themselves.

In addition to categorizing the type of protest, we also want to test the effects of organization types on labor activism (H2). We also coded each entry of the CAiT dataset by the type of organization (if any) responsible for organizing the collective action: employee resource groups (ERGs), employee activism groups (EAGs), and unions. Employee activism groups, or EAGs as we term them, resemble ERGs in that they may be formed around a specific identity group or a specific issue. However, they are not organized or sanctioned by management. Often, these groups emerge after individuals belonging to no particular group organize an action and decide to solidify their collaboration. The group Google Walkout for Real Change came together this way. Although these employee activism groups are relatively informal and have no official membership, some have a logo, a website, Medium, or a GitHub page where they can post open letters and social media accounts.[7] In many cases, these groups are sustained over time and have a track record for organizing multiple protests or campaigns.

Unions are distinguished from ERGs and EAGs on two dimensions. First, firm-based unions typically involve a national union. After the initial surge of employee activism in the industry, a few national unions started to see the tech sector as a viable organizing target. A notable feature of having a national union involved is that they bring external expertise to organizing that professional employees in the tech industry otherwise lack. Second, and more importantly, organizing a union requires formal organizational capacities, such as an organizing committee, dedicated weekly meetings and goal-setting, and usually the involvement of the National Labor Relations Board (NLRB). Building unions require extensive time, energy, and dedication from workers. We expect that these groups will be committed to building lasting organizations that can independently represent workers' interests.

To extract information about the type of organization responsible for organizing an action, we examined news articles about each action to search for evidence that an organization was involved in organizing the action. Where the protest strategy was to release an open letter, it was easy to determine whether an organization was involved because the organization would sign the open letter with their organization's name. For example, when Microsoft employees protested against their employer's contract with the Immigration and Customs Enforcement (ICE) agency, they signed their open letter as "MSWorkers4Good" which had a corresponding X (formerly Twitter) account that posted the letter on their account. This made it clear that the protest action was organized by an EAG. However, not all protests involved an open letter, so we had to rely on the news reporting of the protest to determine if any organization was involved.

---

[7] See Amazon Workers for Climate Justice (2024): https://www.amazonclimatejustice.org/ or Microsoft Workers 4 Good: https://github.com/MSWorkers

We recognize that this likely means underrepresenting the role of EAGs and ERGs in our dataset.

**Methods of analysis**

Using a logistic regression model, we evaluate the effects of social activism and organizational types on the likelihood of labor activism occurring at a company on any given day. The unit of analysis is company-day. Our sample includes all companies with any recorded instances of workplace social activism or labor activism. If a company has no instances of labor activism, they are not included in our sample since our objective is to explain instances of labor activism within tech companies.

We constructed panel data from our archive: for every day $i$ of the study period, for every company $j$, we created a dummy variable $Y_i$ for whether or not labor activism occurred (our dependent variable), as well as four dummy variables $\{S_{i1}, S_{i3}, S_{i6}, S_{i12}\}$ for whether social activism had occurred in the prior N = {1, 3, 6, or 12} months (our independent variable), and four dummy variables $\{L_{i1}, L_{i3}, L_{i6}, L_{i12}\}$ for whether labor activism had occurred in the prior N months (our control). We test multiple lags to see how long it takes social activism to have an impact on labor activism. In addition, we test how different organizational types (EAG, union, or none) affect the likelihood of labor activism. For example, if an instance of activism organized by a union had occurred at this workplace prior to the month in question, we categorized that month as one in which an action was organized by a union in the workplace, represented with the dummy variable $U_i$. Dummy variables were similarly constructed for actions by Employee Activism Groups, $A_i$. In all, there were 100,067 data points in our sample for 56 firms.

Because instances of activism are relatively uncommon at any given company, we chose a logistic probability model, better suited to rare events than a linear probability model (King and

Zheng, 2001). We added firm fixed effects to account for unobserved company-specific characteristics.

$$\text{Model 1: } Y_i = \beta_0 + \beta_1 S_{iN} + \beta_2 L_{iN} + \beta_3 U_{ij} + \beta_4 A_{ij} + \beta_5 FIRM + \epsilon$$

The intercept, $\beta_0$, represents the odds ratio of labor activism happening in a company in a month where no prior activism had happened of either type in the past N months (where N = 1, 3, 6, or 12, depending on the model), at a company where there is no union or EAG. The activism main effects, $\beta_1$ and $\beta_2$, represent the additional likelihood of labor activism given prior social or labor activism in the past N months. The organizational strength main effects, $\beta_3$ and $\beta_4$, represent the additional likelihood of labor activism given the presence of a union or EAG. The firm fixed effects are represented with the vector $\beta_5$: the coefficient in the vector corresponding with the respective firm represents the additional likelihood of labor activism in that firm compared to the reference firm (ActBlue). These coefficients are odds ratios, and therefore if $\beta_1$ is greater than 1, it indicates that the presence of prior social activism in the past N months increases the likelihood of labor activism at a company, and if $\beta_1$ is between 0 and 1, it indicates that prior social activism in the last N months decreases the likelihood of labor activism. Similarly, $\beta_2$ indicates how much more or less likely labor activism is at a company that has had prior instances of labor activism in the last N months. These two coefficients are the main outcomes of interest. $\varepsilon_i$ represents robust standard errors clustered by firm.

## Results

**Descriptive evidence**

Using this basic topology of *social* and *labor* activism, our data analysis illustrates a clear shift in the proportion of social-to-labor activism over the period we examined. Figure 1 presents the split between collective actions categorized as social and labor activism. Figure 2 presents the split as a percentage of total collective actions over the same period, showing a clear shift from actions of the former category to the latter. As Figure 2 shows, the proportion of labor activism has grown from just 10 percent in 2018 to over 75 percent in 2022 and is even higher (20 and 85 percent) when counting actions that are categorized as both social and labor activism. Because our data relies on news articles, the gradual decline in the total number of collective actions observed may not necessarily indicate an overall decrease in tech organizing but a shift away from media-centric tactics (such as walkouts and open letters) in favor of internal or underground labor organizing.[8]

[[Figure 1 here]]

[[Figure 2 here]]

In addition to identifying a clear and gradual shift from social to labor activism in the workplace (Figure 2), it is worthwhile to note that our dataset also illustrates that neither social nor labor activism in the tech industry has been dominated by a single issue or social movement. Figure 3 presents the breakdown of social activism events from 2018 to 2022 by the most prevalent issues across the tech industry. Figure 4 presents the same data for labor activism. For social activism, several issues reappeared over the course of several years– climate change, military contracts, and opposition to the Trump presidency. Some issues are largely temporal. In 2020, BLM-related protests by tech workers contributed more than a third of documented actions but have logged no actions in subsequent years. Labor activism is similarly varied among issues. We see higher and lower levels of labor activism against issues such as sexual harassment, discrimination, and retaliation in some years rather than others, but the issues themselves remain relatively consistent across the industry. Coronavirus, one of the temporal issues, contributes just a small percentage of activism among professional workers. One notable trend in Figure 4 that we will examine in greater detail is the rise in union organizing. Union organizing has increased significantly in terms of the absolute number of events and has been responsible for the multi-year trend in the rise of labor activism in the tech industry.

---

[8] The decline in total collective actions might also demonstrate a shift in interest in the news media from covering collective actions by tech workers. However, labor news coverage has expanded in U.S.-based publications over this period, so we would not expect this to be the case.

[[Figure 3 here]]

[[Figure 4 here]]

When it comes to organizations, our data suggests a strong correlation between the likelihood of labor activism and certain organizational types. We find that a large portion of social activism is organized by EAGs. ERGs only represented one percent of our dataset, so we excluded it from further analysis. Labor activism, not surprisingly, is organized by unions. A large portion of social and labor activism also comes from individual actors who are not organized into a group.

In sum, this analysis illustrates a steady shift from social to labor activism between 2017 and 2022, suggesting a correlation between social and labor activism (H1). It also suggests a correlation between certain organizational types and labor activism (H2). Next, we present the results of our regression to show the effects of social activism and organizational types on the likelihood of labor activism occurring at a company on any given day.

**Regression**

From our logistic regression models, we find a relationship between prior social activism and labor activism (H1). Table 1 shows our descriptive statistics, and Table 2 shows our correlation matrix. Table 3 shows the logged odds ratios of our logistic regression models.

Models 2, 4, 6, and 8 include company fixed effects. Deviance and log-likelihood scores both indicate that the models with the firm fixed effects explain a greater portion of variability. We also include AIC (Akaike Information Criterion) as a goodness of fit measure that penalizes overfitting. AIC measures decrease for models with fixed effects. In our results, we therefore only consider results that are statistically significant *both* with and without company fixed effects. In this way, we ensure that results account for both the greatest amount of variation in our models but also that results cannot be attributed solely to overfitting. All results had a Variance Inflation Factor (VIF) below 7.

[[Table 1 here]]

[[Table 2 here]]

Table 3 demonstrates that, among tech companies where workers organized collective actions, the odds that labor actions would be organized at a company was approximately 1.3 times greater when workers at that company had organized social activism in the prior 6 or 12 months than when workers had not organized social activism in the 12 months prior, supporting H1. Previous social activism seems to, in general, be a stronger predictor of labor activism than previous instances of labor activism. The duration in which labor activism may occur after social activism appears to have an effect: social activism six months and twelve months previously has a positive impact on the likelihood of labor activism occurring in a company. Labor activism six or more months previously appears to decrease the likelihood of labor activism at a company.

[[Table 3 here]]

We believe this variation in the duration may be because labor activism often takes time to organize. Union drives can take several months to several years, and similarly, it could take a long time to build the organizational strength required to conduct a walkout or other work stoppages. During these stretches of time, workers are not only building their numbers by convincing workers to participate but also convincing those workers that participating in a labor action is an effective way to achieve their goals.

We have three suggestions as to why labor activism six or more months prior appears to decrease the likelihood of labor activism. First, two of the labor activism actions in our dataset mark the ratification of a union contract. In our experience, labor organizing often slows down after the culmination of a long, hard contract bargaining process. Second, 20 of the labor actions in our dataset mark the announcement of the recognition of a union (either voluntarily or by NLRB election). It can take months or years to bargain and ratify a new contract, and while it is common for workers to take labor actions during a long contract campaign, it is possible that collective actions during contract campaigns aimed at pressuring management to give in to demands at the bargaining table do not begin until more than one year of bargaining has occurred, due to hopes that the bargaining process itself can achieve the aims that collective action has been used for in the past. For example, the New York Times Tech Guild went public in April, 2021, held a walkout in August, 2021, and was recognized via NLRB election in March, 2022. We have no subsequent instances of activism of any kind by tech workers in this workplace in our dataset, which ends nine months later, in December, 2022, although they held another walkout in October, 2023, to protest return-to-office policies. This could also be due to our dataset: collective actions that are commonly used during the bargaining period, like "sticker up" campaigns, are not frequently reported on in the media. Third, our dataset is reliant on news reports which may not capture the full picture. News interest in labor-related activities is more likely concentrated in the immediate period after the start of a union drive or election (which make up 29.7 percent of our labor actions in the dataset) and may drop off over time.

We find significant positive effects on the odds that workers would engage in labor activism at a given company when prior actions were organized by a union but not when they were organized by an employee activism group (EAG). It is not surprising and, in fact, could be seen as tautological, that labor unions would increase participation in labor activism given that their function is to organize collective labor actions by workers. We find support for only part of H2. The presence of organized groups within a company by itself does not increase the odds of later labor activism. Instead it appears that these groups must be organized with the intention of conducting further labor actions.

**The case of Google**

Above, we systematically demonstrate a relationship between social activism and labor activism. To illustrate, in practice, how participation in social activism creates the necessary conditions for labor activism to occur, we examine a sequence of social and labor activism at Google. We selected Google as our case study because protests by Google employees are well documented in the media. Google employees who have participated in labor activism have also been interviewed extensively by the media. To conduct this case study, we analyzed all collective actions in our dataset involving Google employees and used process tracing to evaluate how events from one event influenced future events. Recognizing that not all Google protests are related, we grouped collective actions by Google employees by the issues at the heart of the actions. We found that many instances of labor activism took place at the end of a time series of actions that began with social activism.

For this case study, we examine a series of collective actions in 2019, regarding Google's contracts with US Customs and Border Patrol (CBP), Immigration and Customs Enforcement

(ICE), and the Office of Refugee Resettlement (ORR).[9] That summer, an EAG calling themselves "No GCP for CBP" published an open letter/petition on Medium demanding that Google pledge not to support CBP, ICE, or ORR by allowing them to use Google's cloud infrastructure services (GCP)—we classified this action as social activism. Google worker Rebecca Rivers helped organize this letter, which stated in part:

> *In working with CBP, ICE, or ORR, Google would be trading its integrity for a bit of profit, and joining a shameful lineage. We have only to look to IBM's role working with the Nazis during the Holocaust to understand the role that technology can play in automating mass atrocity. History is clear: the time to say NO is now. We refuse to be complicit. It is unconscionable that Google, or any other tech company, would support agencies engaged in caging and torturing vulnerable people.*

Open letters such as this one that leveraged a strong sense of moral justice were typical among instances of social activism. A few workers who had signed this open letter, including Rivers and Sophie Waldman, later discovered that Google was, in fact, already working with CBP, and shared this information with coworkers. On November 12th, 2019, Rivers and another Google worker, Laurence Berland, were placed on involuntary leave from Google. Berland had participated in a different instance of social activism, protesting Google's participation in San Francisco's pride parade despite the company's unwillingness to change Youtube's content moderation policies which, at the time, allowed floods of anti-LGBTQ+ comments on the platform and led to harassment of queer creators.

---

[9] On January 30th, 2017, over 2,000 Google workers and members of management staged a walkout (from offices all over the world) in protest of then-President Trump's executive order known as the "muslim ban." The walkout was organized by workers but supported by Google. Google co-founder Sergey Brin and CEO Sundar Pichai joined the walkout at the Mountain View, CA campus and spoke to the crowd. However, their participation in protesting against anti-immigration policy would end here.

On November 19th, 2019, about 20 Google workers accompanied Berland into a meeting with management in protest of what they saw as retaliation against Berland's and Rivers' organizing. On November 22nd, 2019, around 200 Google workers showed up to a protest in front of Google's San Francisco office, demanding Rivers' and Berland's reinstatement. Because these protests were directly concerned about disciplinary actions taken on coworkers, we categorized both as instances of labor activism. Notably, no particular worker organization was affiliated to these labor protests.

On November 25th, 2019, three days before Thanksgiving, Berland, Rivers, Waldman, and a fourth organizer named Paul Duke were fired. Google's stated reason was that they had violated company policies by accessing company documents that were "outside the scope of their jobs." On December 3rd, 2019, the "Thanksgiving Four" posted an open letter on Medium, and announced that they were filing an Unfair Labor Practice charge (ULP) with the National Labor Relations Board (NLRB)[10]. The open letter states:

> *Google wants to send a message to everyone: if you dare to engage in protected labor organizing, you will be punished. […] Our answer is the same for everyone working across the tech industry: Now is the time to organize, to join with your colleagues, and hold the bosses accountable! Until we all come together in solidarity, for our workplace, for our communities, and for our world, nothing will change. But every one of us knows what we need and what the world deserves, and together we can make a difference.*

In stark contrast to the moralistic language that Google employees used to oppose the company's contract with CBP, this letter adopts the language of labor organizing and demonstrates class consciousness.

---

[10] The four were later joined by Kathryn Spiers, who was fired for a different protest action at the company, making them the "Thanksgiving Five."

This 2019 case study of social activism at Google extends Fantasia's cultures of solidarity argument—that worker's capacity and drive to organize emerges out of conflict with management—to professional tech workers. It shows that tech workers, motivated by their unique social justice-centered professional identity, may participate in social activism when they see their company pursue profit at the expense of its socially-driven mission. However, when they seek resolutions to their problems and their efforts are frustrated/or retaliated against, they drop their social-justice-centered professional identities and come to recognize themselves as workers, thus pushing them to use the tools of class struggle. We see such a transformation as the basis on which sustained labor activism is possible.

## Discussion and conclusion

Professionals are not conventionally associated with labor activism. Tech workers even less so. However, our article provides evidence of tech workers, one of the highest-paid types of professionals, participating in labor organizing across the sector. This article asks what is behind the rise of labor activism among tech workers. To answer this question, we examined all documented instances of collective action by U.S.-based professional tech workers from 2014 to 2022. Our analysis comprises three parts. First, descriptive statistics show that labor activism was preceded by a surge in social activism in the workplace—a distinct form of workplace activism that concerns issues in which participating employees are not directly impacted by the issue in question. Second, using a logistic regression model and company fixed effects to test the relationship between the two types of workplace activism, we find that U.S.-based tech workers are more likely to engage in labor activism if they have a history of social activism at their company. Finally, to explore the mechanism in which one leads to the other, we conducted a case study of the transformation of social activism to labor activism at Google. We find that for tech workers, participation in political, cultural, or identity-based struggles generates conflict with management and exposes the cleavage between workers and management, thereby creating the necessary conditions for labor activism.

With these findings, we contribute to the literature in two ways. Existing literature suggests that professionals, from their advantaged position, have neither the need nor the interest to challenge existing systems of power in the workplace. Alongside recent research on teachers and nurses (Weiner 2012; Dube, Kaplan, and Thompson 2016; Gastón 2022), our first contribution is to add the case of tech professionals to the growing evidence that professionals do, in fact, engage in labor activism. Second, we offer a new perspective on professional tech workers. The literature on tech workers has focused on the culture of the tech industry and, more recently, the culture of tech work, which describes tech workers as individualistic and aligned with management, making them unlikely candidates for labor activism (Dorschel 2022a; Widder and Nafus 2023). What little there is on workplace activism in the tech industry has, as a result, theorized tech worker protests as a product of that culture (Tarnoff 2020; Boag et al. 2022; Luka and Tan 2022). While recognizing these features of tech work, this article draws attention to the rise of labor activism and how tech workers, through participating in social activism, developed class consciousness and a worker identity, thereby laying the foundation for labor activism.

Fantasia (1988) demonstrated how professional and worker identities can be synergistic rather than antagonistic and can encourage previously reluctant professionals to engage in labor activism. In one of his case studies, nurses developed a worker identity partly in reaction to management's anti-union campaigns. Professional identities bolstered nurses' demands by positioning them as authorities on patient care. Management's pushback on demands became seen as a misalignment of professional values as well as a conflict along class lines. We documented a similar phenomenon among tech workers. In the mid-2010s, many tech workers believed that working towards the greater social good was a shared professional value. When management responded to social activism demands with retaliation, however, it became clear to tech workers that their values were not shared with their bosses. In this way, social activism became an instigating force for later labor activism. We see this mechanism play out in detail in our case study of Google employees.

One future research avenue is to unpack the role that employee organizations play in the transition from social to labor activism. While the presence of a union does have a strong effect on the likelihood of labor activism, other employee organizations—in particular, employee activist groups—did not affect the likelihood of labor activism, according to our regression. We recognize that EAGs do not all operate in the same way and can thus represent vastly different degrees of organizational strength. For example, Amazon's EAG—Amazon Employees for Climate Justice—was formed in 2019 and has sustained itself as a group to this date. Some EAGs, such as Google Walkout For Real Change, even laid the groundwork for involving national unions. By contrast, Facebook employee's EAG, Workers4Workers, was only active for one year. This variation in EAGs suggests that future work can be done to dissect the relationship between social activism, organizational strength, and labor activism.

The biggest limitation in our analysis is the reliance on news articles, which give us insight only into public events that journalists find newsworthy. We can think of many instances of protest action of the type archived in our dataset that never garnered press, such as open letters that were never leaked. Past studies of social movements have relied heavily on news archives precisely because it is so difficult to attain a high-level view of a social movement across multiple geographies and organizations any other way (Earl et al. 2004). Recently, social movement scholars have been able to develop unique methods to study social movements relying on social media data, most prominently Twitter (Wood and Goldstein 2023). However, our experience cataloging workplace activism in the U.S. tech industry suggests that workplace protests are far less likely to appear on social media than street protests because workers are more likely to promote these protests through company-level communications such as email lists or private messaging channels.

Another shortcoming is that collective actions supported by informal intra-industry labor networks like the Tech Workers Coalition (TWC), Jobs with Justice[11], Collective Action in Tech, and the Emergency Workers Organizing Committee (EWOC) are not explored. Because our dataset relies on news reports that often gloss over the role of such third-party organizations, we don't have accurate data to theorize their exact role. Yet, as previous literature has shown, the effects of these intra-industry organizational forms should not be underestimated for their ability to make an impact on entire industries through coalition-building and the formation of new organizations and shared resources (Dixon, Tope, and Van Dyke 2008; Wang and Soule 2012).

Finally, it is important to acknowledge that although our data indicates a significant increase in labor activism over the past five years, tech workplace activism is still in its early stages. Our data suffers from right-hand censoring, so we must acknowledge that it is too early to know whether what we've observed is an aberration from a long-term trend or a sign of a lasting change in the trend of professional labor organizing itself. In particular, while our descriptive analysis shows that the wave of tech labor organizing began several years before the tech downturn and austerity era, which began in late 2022, our dataset, which ends in 2022, does not measure the effects of this era on labor activism in the tech industry. Moreover, the most significant rounds of layoffs coming from the tech downturn took place at the end of 2022 and 2023. The story of labor activism in tech is still an unfolding one.

---

[11] See: Boston Tech Workers for Justice. "About." Boston Tech Workers for Justice. http://www.btwj.org/

Despite these limitations, this article provides new evidence for why professional workers may, as an outcome of social activism, begin to participate in labor activism. We see this development as a positive one. In an increasingly powerful and unaccountable tech industry, we believe that this work of labor organizing is precisely the type of work that can create a lasting impact on democratizing the industry and the future of work.

**Figures**

Figure 1. Collective actions by social versus labor activism

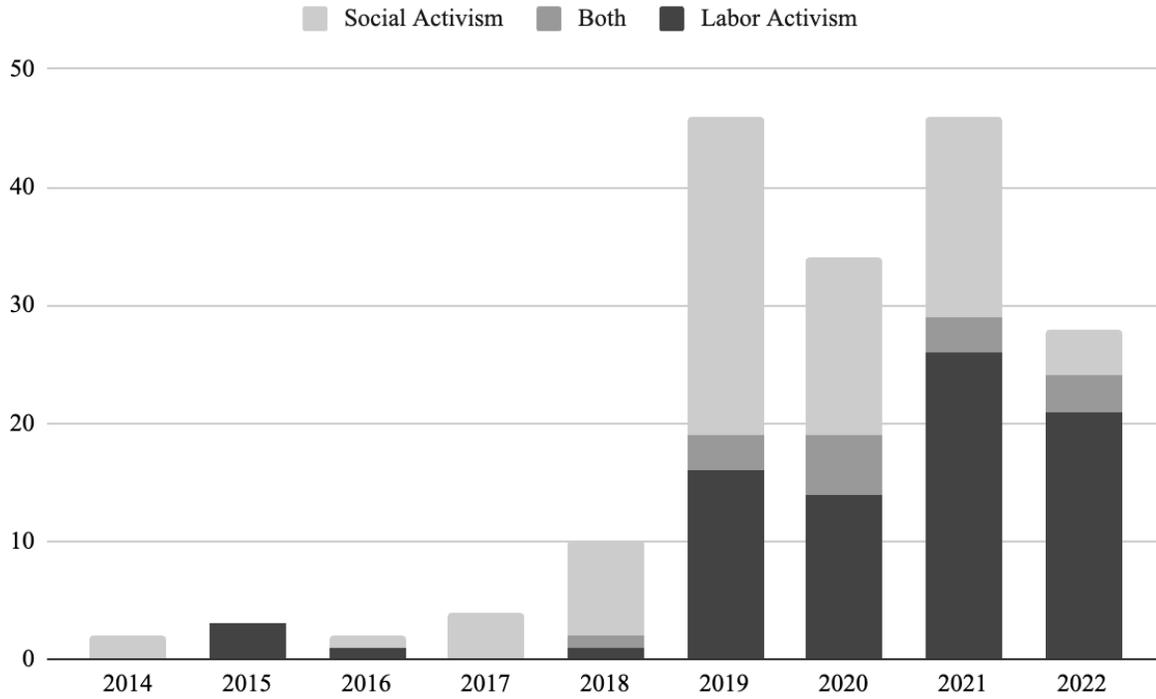

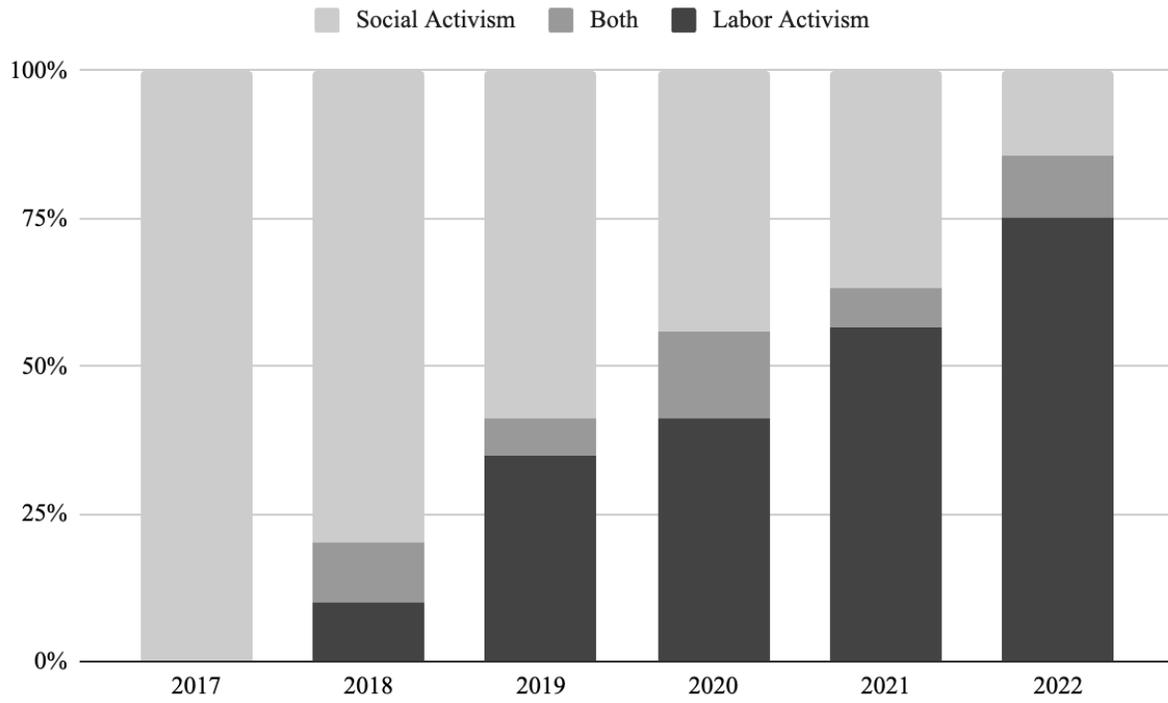

Figure 2. Proportion of social versus labor activism by year[12]

---

[12] We excluded the years 2014, 2015, and 2016 from Figure 3 because each of those years had fewer than three collective actions.

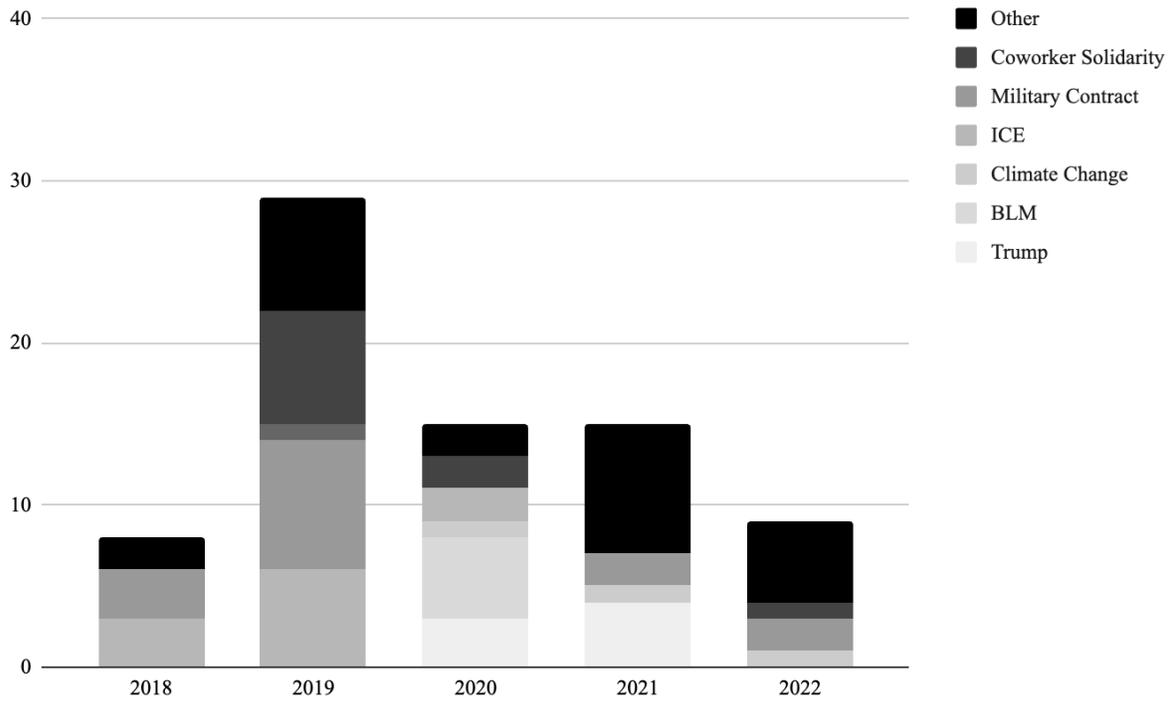

Figure 3. Social activism by issue (2018–2022) [13]

---

[13] "Coworker solidarity" protests are not directly tied to professional workers' own working conditions but are shows of solidarity with other types of workers at the company.

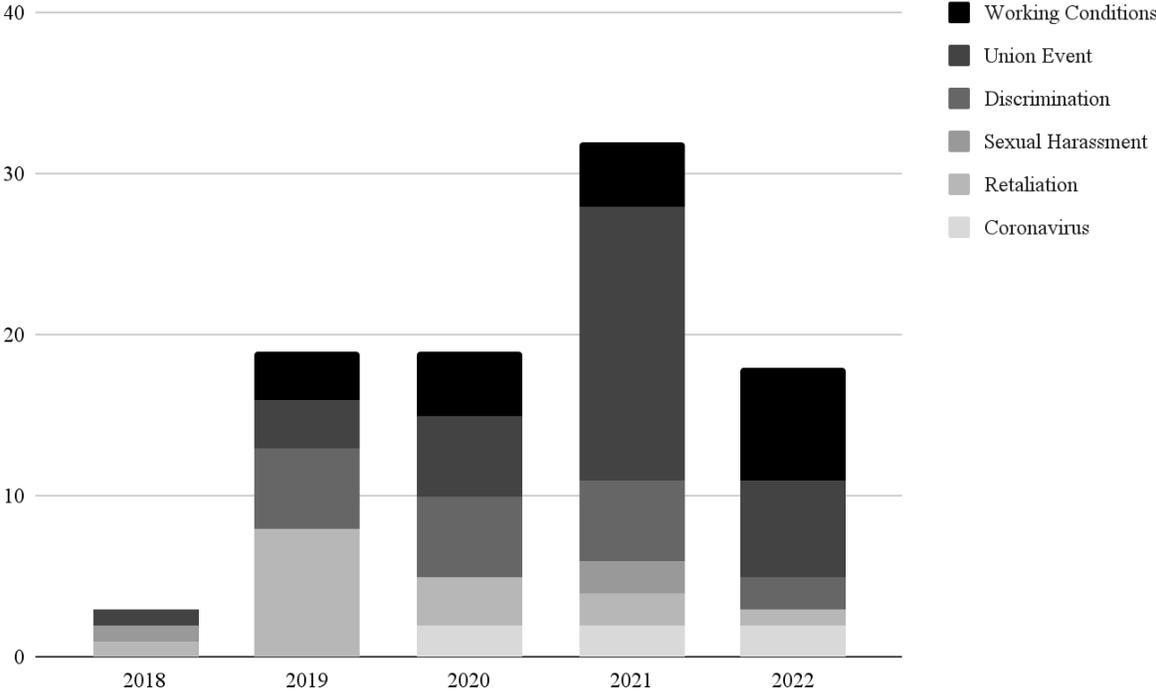
Figure 4. Labor activism by issue (2018–2022)

**Tables**

Table 1. Descriptive Statistics, 2014-2022

| **Statistic** | **Mean** | **St. Dev.** | **Min** | **Max** |
|---|---|---|---|---|
| Social Activism (1 mo) | 0.0282 | 0.2024 | 0 | 4 |
| Social Activism (3 mo) | 0.0855 | 0.4084 | 0 | 6 |
| Social Activism (6 mo) | 0.1692 | 0.6775 | 0 | 9 |
| Social Activism (1 y) | 0.3294 | 1.1869 | 0 | 12 |
| Labor Activism (1 mo) | 0.0276 | 0.3478 | 0 | 4 |
| Labor Activism (3 mo) | 0.0827 | 0.5516 | 0 | 5 |
| Labor Activism (6 mo) | 0.1612 | 0.9001 | 0 | 7 |
| Labor Activism (1 y) | 0.3016 | 0.3649 | 0 | 9 |
| Union | 0.1582 | 0.3071 | 0 | 1 |
| EAG | 0.1054 | 0.1010 | 0 | 1 |

Note: n = 100,067 for all variables

Table 2. Correlation Matrix, 2014-2022

| 1 | 2 | 3 | 4 | 5 | 6 | 7 | 8 | 9 | 10 |
|---|---|---|---|---|---|---|---|---|---|
| 1 | - | - | - | - | - | - | - | - | - |
| 0.66 | 1 | - | - | - | - | - | - | - | - |
| 0.57 | 0.83 | 1 | - | - | - | - | - | - | - |
| 0.48 | 0.73 | 0.88 | 1 | - | - | - | - | - | - |
| 0.28 | 0.28 | 0.30 | 0.29 | 1 | - | - | - | - | - |
| 0.29 | 0.45 | 0.47 | 0.47 | 0.64 | 1 | - | - | - | - |
| 0.30 | 0.47 | 0.58 | 0.60 | 0.52 | 0.81 | 1 | - | - | - |
| 0.29 | 0.46 | 0.58 | 0.69 | 0.41 | 0.67 | 0.85 | 1 | - | - |
| 0.07 | 0.11 | 0.14 | 0.17 | 0.20 | 0.32 | 0.41 | 0.50 | 1 | - |
| 0.27 | 0.41 | 0.49 | 0.58 | 0.16 | 0.26 | 0.34 | 0.40 | 0.12 | 1 |

1 = Social Activism (1 mo), 2 = Social Activism (3 mo), 3 = Social Activism (6 mo), 4 = Social Activism (1 y), 5 = Labor Activism (1 mo), 6 = Labor Activism (3 mo), 7 = Labor Activism (6 mo), 8 = Labor Activism (1 y), 9 = Union (Binary), 10 = EAG (Binary)

Table 3. Logistic Regression of Labor Actions Following Social Activism, 2014-2022

|  | 1 Month | | 3 Months | | 6 Months | | 1 Year | |
| --- | --- | --- | --- | --- | --- | --- | --- | --- |
|  | Model 1 | Model 2 | Model 3 | Model 4 | Model 5 | Model 6 | Model 7 | Model 8 |
| Social Activism (1 mo) | 1.47 (0.22) | 1.20 (0.24) | | | | | | |
| Social Activism (3 mo) | | | 1.30* (0.12) | 1.13 (0.13) | | | | |
| Social Activism (6 mo) | | | | | 1.39** (0.11) | 1.31* (0.11) | | |
| Social Activism (1 y) | | | | | | | 1.37*** (0.09) | 1.29** (0.08 |
| Labor Activism (1 mo) | 1.65* (0.24) | 1.25 (0.26) | | | | | | |
| Labor Activism (3 mo) | | | 1.18 (0.16) | 0.91 (0.17) | | | | |
| Labor Activism (6 mo) | | | | | 0.90 (0.15) | 0.69** (0.14) | | |
| Labor Activism (1 y) | | | | | | | 0.78* (0.12) | 0.44*** (0.18) |
| Union | 6.52*** (0.24) | 8.01*** (0.39) | 6.37*** (0.25) | 8.29*** (0.40) | 6.76*** (0.27) | 10.32*** (0.41) | 8.26*** (0.27) | 33.43*** (0.51) |
| EAG | 3.32*** (0.24) | 0.91 (0.55) | 2.96*** (0.25) | 0.95 (0.56) | 2.43** (0.28) | 1.02 (0.58) | 2.11* (0.31) | 2.13 (0.67) |
| Firm Fixed Effects | NO | YES | NO | YES | NO | YES | NO | YES |
| AIC | 1323.87 | 1366.31 | 1325.28 | 1367.74 | 1318.34 | 1358.34 | 1318.01 | 1325.44 |
| Log Likelihood | -656.93 | -623.15 | -657.64 | -623.87 | -654.17 | -619.17 | -654.01 | -602.72 |
| Deviance | 1313.87 | 1246.31 | 1315.28 | 1274.74 | 1308.34 | 1238.34 | 1308.01 | 1205.44 |
| No. Obs. | 100,067 | 100,067 | 100,067 | 100,067 | 100,067 | 100,067 | 100,067 | 100,067 |

*** p <0.001, ** p <0.01, * p < 0.05